\documentclass{aa}  

\usepackage{graphicx}
\usepackage{txfonts}
\usepackage{natbib} 
\bibpunct{(}{)}{;}{a}{}{,} 

\usepackage{verbatim} 
\usepackage{hyperref}   
\hypersetup{colorlinks=true,linkcolor=blue,citecolor=blue,filecolor=blue,urlcolor=blue}

\begin{document}

   \title{Constraining the helium-to-metal enrichment ratio $\Delta Y/\Delta Z$ from main sequence binary stars}

   \subtitle{Theoretical analysis of the accuracy and precision of the age and helium abundance estimates}
  \author{G. Valle \inst{1, 2}, M. Dell'Omodarme \inst{1}, P.G. Prada Moroni
        \inst{1,2}, S. Degl'Innocenti \inst{1,2} 
}
\titlerunning{$\Delta Y /\Delta Z$ form binary stars}
\authorrunning{Valle, G. et al.}

\institute{
        Dipartimento di Fisica "Enrico Fermi'',
        Universit\`a di Pisa, Largo Pontecorvo 3, I-56127, Pisa, Italy
        \and
        INFN,
        Sezione di Pisa, Largo Pontecorvo 3, I-56127, Pisa, Italy
}

   \offprints{G. Valle, valle@df.unipi.it}

   \date{Received ; accepted }

  \abstract
{}
{We investigated the theoretical possibility of accurately determining the helium-to-metal enrichment ratio $\Delta Y/\Delta Z$ from precise observations of double lined eclipsing binary systems. }
{Using Monte Carlo simulations, we drew synthetic binary systems with masses between 0.85 and 1.00 $M_{\sun}$ from a grid of stellar models. Both stars were sampled from a grid with $\Delta Y/\Delta Z = 2.0$, with primary star at 80\% of its main-sequence evolution. Subsequently, a broader grid with $\Delta Y/\Delta Z$ from 1.0 to 3.0 was used in the fitting process. To account for observational uncertainties, two scenarios were explored: S1 with realistic uncertainties of 100 K in temperature and 0.1 dex in [Fe/H], and S2 with halved uncertainties. We repeated the simulation at two baseline metallicities: [Fe/H] = 0.0 and $-0.3$.
}
{The posterior distributions of $\Delta Y/\Delta Z$ revealed significant biases. The distributions were severely biased towards the
edge of the allowable range in the S1 errors scenario. The situation only marginally improved when considering
the S2 scenario. The effect is due to the impact of changing $\Delta Y/\Delta Z$ in the stellar effective temperature and its interplay with [Fe/H] observational error, and it is therefore not restricted to the specific fitting method. 
Despite the presence of these systematic discrepancies, the age of the systems were recovered unbiased with 10\% precision.
}
{ Our findings indicate that the observational uncertainty
in effective temperature and metallicity significantly hinders the accurate determination of the $\Delta Y/\Delta Z$ parameter from main sequence binary systems. } 


   \maketitle

\section{Introduction}\label{sec:intro}

While stellar model predictions have significantly improved in recent decades, they still face shortcomings in their modelling of some physical phenomena, such as convective transport \citep[see][for an  introduction]{Viallet2015} or the effects of microscopic diffusion and of the mechanisms competing with it \citep[see e.g.][]{Moedas2022}.
In this context, binary systems are often adopted to calibrate free parameters in stellar model computations. This is particularly true following the improvement in the radii and masses precision for stars in binary system (thanks to satellite missions such as Kepler and TESS \citealt{Borucki2010, Ricker2015}), alongside with progressions in the methods of estimating the stellar effective temperatures \citep{Miller2020}. Nowadays a precision in the measured parameters below 1\% is not uncommon and several binary systems have masses and radii measured at 0.1\% precision. 

In addition to their sensitivity to the underlying physics, stellar models are also influenced by the assumed chemical composition, particularly the abundance of helium ($Y$) and the total metallicity ($Z$). 
While surface metallicity can be measured from absorption lines of several metals in the stellar spectra, determining the helium abundance for the majority of stars is not possible due to the absence of observable helium lines in stars cooler than approximately 20,000 K. 
From a theoretical perspective, this necessitates the assumption of an initial helium abundance for stellar evolution computations.
The conventional approach involves adopting a linear relationship between the  helium abundance and metallicity: $Y = Y_p+\frac{\Delta Y}{\Delta Z} Z$, where $Y_p$ is the primordial helium abundance produced in the Big Bang nucleosyntehsis and $\Delta Y/\Delta Z$ is the helium-to-metal enrichment ratio.  

The determination of the value of $\Delta Y/\Delta Z$ has received significant attention in the literature. 
Several methods have been employed to constrain its value, including comparing isochrones and observations in the HR diagram \citep[e.g.][]{pagel98, Casagrande2007,  gennaro10, Tognelli2021},
fitting stellar tracks on a survey of nearby field stars \citep{Valcarce2013}, 
constructing a  standard solar model that reproduces the Sun's present luminosity, radius, and
surface $Z/X$ ratio \citep[e.g][]{gennaro10, Serenelli2010, Tognelli2011, Valcarce2012}, 
calibrating models with evolved stars (horizontal branch and red giant stars, 
\citealt{Renzini1994, Marino2014,  Valcarce2016}),
using galactic and extragalactic H II regions \citep[e.g.][]{Peimbert1974, Pagel1992, Chiappini1994, Peimbert2000, Fukugita2006, Mendez2020}
or planetary nebula \citep{Dodorico1976, Chiappini1994, Peimbert1980, Maciel2001}. 
The results from these diverse approaches exhibit a large variability and potential systematic uncertainties. 
For instance,  standard solar models suggest $\Delta Y/\Delta Z \lesssim 1$, whereas evolved stars favour $\Delta Y/\Delta Z \in [2, 3]$, 
and HR diagram comparisons typically indicate values within the [1, 3] range.  

Another possibility to constrain the helium-to-metal enrichment ratio comes from detached double-lined eclipsing binaries (EBs). These systems offer the opportunity to precisely determine the masses, radii, and effective temperatures of their stellar components. Since these stars are presumed to share a common age and initial chemical composition, they serve as ideal experimental environment for stellar evolution models calibration. By comparing these models to observed properties of EBs, it is possible to gain deeper insights into fundamental stellar processes, such as the treatment of convection and diffusion \citep[see among many][]{Andersen1991, Torres2010,  Nataf2012b, TZFor, Claret2017, Valle2023b}.
Early investigations \citep{Ribas2000, Fernandes2012} suggested a value of 
$\Delta Y/\Delta Z = 2 \pm 1$, with some systems exhibiting higher values.
However, the 
primordial helium abundance inferred from these studies was $Y_p = 0.225$,
significantly lower than the primordial abundance estimated from Big Bang nucleosynthesis.
The increasing precision in mass and radius measurements from binary systems, nowadays often reaching 0.1\% accuracy, suggests an unprecedented opportunity to further refine the estimation of 
$\Delta Y/\Delta Z$. 

In this paper, we theoretically explore this possibility through the creation of artificial stars generated from a grid of computed stellar models. By quantifying the statistical and systematic uncertainties affecting the $\Delta Y/\Delta Z$ estimates in a fully controlled environment, we established the highest achievable precision from EBs. 
The investigated systems were carefully selected due to the complexity of estimating the helium-to-metal enrichment ratio. This estimation involves simultaneously fitting additional stellar model parameters, such as age, convective core overshooting efficiency, and mixing-length value
\citep[see e.g.][]{Feiden2015}. 
These parameters significantly impact stellar model calculations of solar-like stars, introducing degeneracies in the fit. Therefore, attempts to fit the $\Delta Y/\Delta Z$ value from observations should rely on carefully chosen EBs. For instance, limiting the analysis to main sequence (MS) stars without convective cores allows for the neglect of the convective core overshooting efficiency, effectively eliminating a nuisance parameter.
 
The structure of the paper is as follows. In Sect.~\ref{sec:method}, we 
discuss the grid, the sampling scheme and the fitting method used in the estimation process. 
The  results are
presented in Sect.~\ref{sec:results}, with an analysis of the metallicity effect.
Some concluding remarks can be found in Sect.~\ref{sec:conclusions}.

\section{Methods}\label{sec:method}
\subsection{Stellar models grid}

The grid of stellar evolutionary models was calculated for the mass range 0.85 to 1.00 $M_{\sun}$, spanning the evolutionary stages from the pre-main sequence to the onset of the RGB.
The initial metallicity [Fe/H] was varied from $-0.5$ dex to 0.2 dex with
a step of 0.01 dex. 
We adopted the solar heavy-element mixture by \citet{AGSS09}. 
For each metallicity, we considered a range of initial helium abundances based on the commonly used linear relation $Y = Y_p+\frac{\Delta Y}{\Delta Z} Z$
with the primordial helium abundance  $Y_p = 0.2471$ from \citet{Planck2020}.
The helium-to-metal enrichment ratio $\Delta Y/\Delta Z$ was varied
from 1.0 to 3.0 with a step of 0.1. 

Models were computed with the FRANEC code, in the same
configuration as was adopted to compute the Pisa Stellar
Evolution Data Base\footnote{\url{http://astro.df.unipi.it/stellar-models/}} 
for low-mass stars \citep{database2012}. The only difference with respect to those models is that  in the present research the outer boundary conditions  were set by the \citet{Vernazza1981} solar semi-empirical $T(\tau)$, which approximate well results obtained using the hydro-calibrated $T(\tau)$ \citep{Salaris2015, Salaris2018}.
The models were computed
assuming the solar-scaled mixing-length parameter $\alpha_{\rm
	ml} = 2.02$.
Atomic diffusion was included adopting the coefficients given by
\citet{thoul94} for gravitational settling and thermal diffusion. 
To prevent extreme variations in the surface chemical abundances 
the diffusion velocities were
multiplied by a suppression parabolic factor that takes a value of 1 for 99\% of the mass of the structure and 0 at the base of the atmosphere \citep{Chaboyer2001}.

Raw stellar evolutionary tracks were reduced to a set of tracks with the same number of homologous points according to the evolutionary phase.
Details about the reduction procedure are reported in the Appendix of \citet{incertezze1}.  Given the accuracy in the observational radius data,  a linear interpolation in time was performed for every reduced track in order to ensure that the separation in radius between consecutive track points in the $10 \sigma$ range from the observational targets was less than one-quarter of the adopted radius uncertainty.

\subsection{Sampling scheme}\label{sec:sampling}

Artificial binary systems were constructed as follows.
\begin{itemize}
	\item Primary stars were selected from the grid of models with mass values in the set $M_p = 0.85, 0.90, 0.95, 1.00$ $M_{\sun}$, metallicities [Fe/H] = 0.0, and $-0.3$, with $\Delta Y/\Delta Z = 2.0$. 
	The stars were selected when they reached 80\% of their MS lifetime, as this is when the effect of the helium abundance difference is most pronounced \citep{binary}. The lower mass boundary was chosen in a way that every star could reach the evolutionary sampling point within the age of the Universe.
	The higher boundary was determined such that convective core overshooting could be safely disregarded, because stars in the chosen mass range burn hydrogen in a radiative core.
	\item Each primary star was paired with a secondary within the same masses set, limited to the primary's mass value, and having the same chemical composition and age as the primary star. This selection yielded  10 systems for each [Fe/H] value.
	\item The observational constraints taken into account were effective temperature ($T_{\rm eff}$), metallicity [Fe/H], mass, and radius. We also tested a different scenario adopting stellar log luminosity instead of effective temperature as observational constraint, obtaining a noticeable agreement in the results. We therefore focus only on the former scenario. 
	Two sets of observational errors were considered: the first set (S1) represents a realistically achievable precision of 100 K in $T_{\rm eff}$, 0.1 dex in [Fe/H], 0.5\% in radius and mass. 
	While it is not uncommon to find studies in the literature reporting an accuracy of a few tenths of a K in $T_{\rm eff}$, direct comparisons of results obtained by different authors for the same stars often reveal discrepancies exceeding 100 K \citep[see e.g.][]{Ramirez2005,Schmidt2016}. 	
	The second set (S2) assumes over-optimistically precise measurements of 50 K in $T_{\rm eff}$, 0.05 dex in [Fe/H]  \citep[see e.g.][]{Yu2023, Hegedus2023}, 0.5\% in radius and mass. 
 The selected system observables were subjected to Gaussian random perturbations to simulate observation uncertainties, while taking into account the correlation structure among the observational data for the two stars. 
	Following \citet{TZFor, Valle2023b}, we assumed a correlation of 0.95 between	the primary and secondary effective temperatures, and 0.95
	between the metallicities of the two stars. Regarding mass and radius correlations, the high precision of the estimates means that these parameters are of no importance,
	For each binary system $N = 5\,000$ perturbed systems were generated.
	\item Each perturbed system underwent parameter estimation (as explained in Sect.~\ref{sec:fit-method}), assuming identical chemical composition and age for both the primary and secondary stars. Only solutions providing a satisfactory fit
	(determined by their $\chi^2$ values) were retained in the final sample. From these samples, marginalized posterior distributions of the age and helium-to-metal enrichment ratio were obtained.
\end{itemize}

\subsection{Fitting technique}\label{sec:fit-method}

The analysis is conducted adopting the SCEPtER pipeline\footnote{Publicly available on CRAN: \url{http://CRAN.R-project.org/package=SCEPtER}, \url{http://CRAN.R-project.org/package=SCEPtERbinary}}, a well-tested technique for fitting single and binary systems
\citep[e.g.][]{scepter1,eta, binary, TZFor, Valle2023a}. 
The pipeline estimates  the parameters of interest (i.e. the system age and its initial chemical abundances) adopting a grid maximum likelihood  approach.

The method we use is explained in detail in \citet{binary}; here, we provide only a brief summary for convenience. For every $j$-th point in the fitting grid of precomputed stellar models, a likelihood estimate is obtained for both stars:
\begin{equation}
        {{\cal L}^{1,2}}_j = \left( \prod_{i=1}^n \frac{1}{\sqrt{2 \pi}
                \sigma_i} \right) 
        \times \exp \left( -\frac{\chi^2}{2} \right)
        \label{eq:lik}
        ,\end{equation}
\begin{equation}
        \chi^2 = \sum_{i=1}^n \left( \frac{o_i -
                g_i^j}{\sigma_i} \right)^2
        \label{eq:chi2},
\end{equation}
where $o_i$ are the $n$ observational constraints, $g_i^j$ are the $j$-th grid point corresponding values, and $\sigma_i$ are the observational uncertainties.

The joint likelihood of the system is then computed as the product of the single star likelihood functions.  Information about 
effective temperature and mass ratios was adopted as multiplicative Gaussian priors. However, we directly verified that their impact in the results was negligible, because the correlations in the Monte Carlo samples (Sect.~\ref{sec:sampling})  translate in correlations between the fitted parameters (see Appendix~\ref{app:corr}).
It is possible to obtain estimates both for the individual components and for 
the  whole system. In the former case, the fits for the two stars are obtained independently,
while in the latter case the two objects must have a common age (with a tolerance of 1 Myr), identical initial helium abundance, and initial metallicity. Stellar parameter estimates are obtained averaging the corresponding quantity of all the models with likelihood greater than $0.95
\times {\cal L}^{1,2}_{\rm max}$\footnote{
The likelihood threshold is set to improve computational efficiency. This approach introduces negligible differences compared to using a weighted mean where each model's weight reflects its likelihood.}.

\section{Results}\label{sec:results}

\subsection{Artificial stars with [Fe/H] = 0.0}

\begin{figure*}
	\centering
	\includegraphics[width=17.5cm,angle=0]{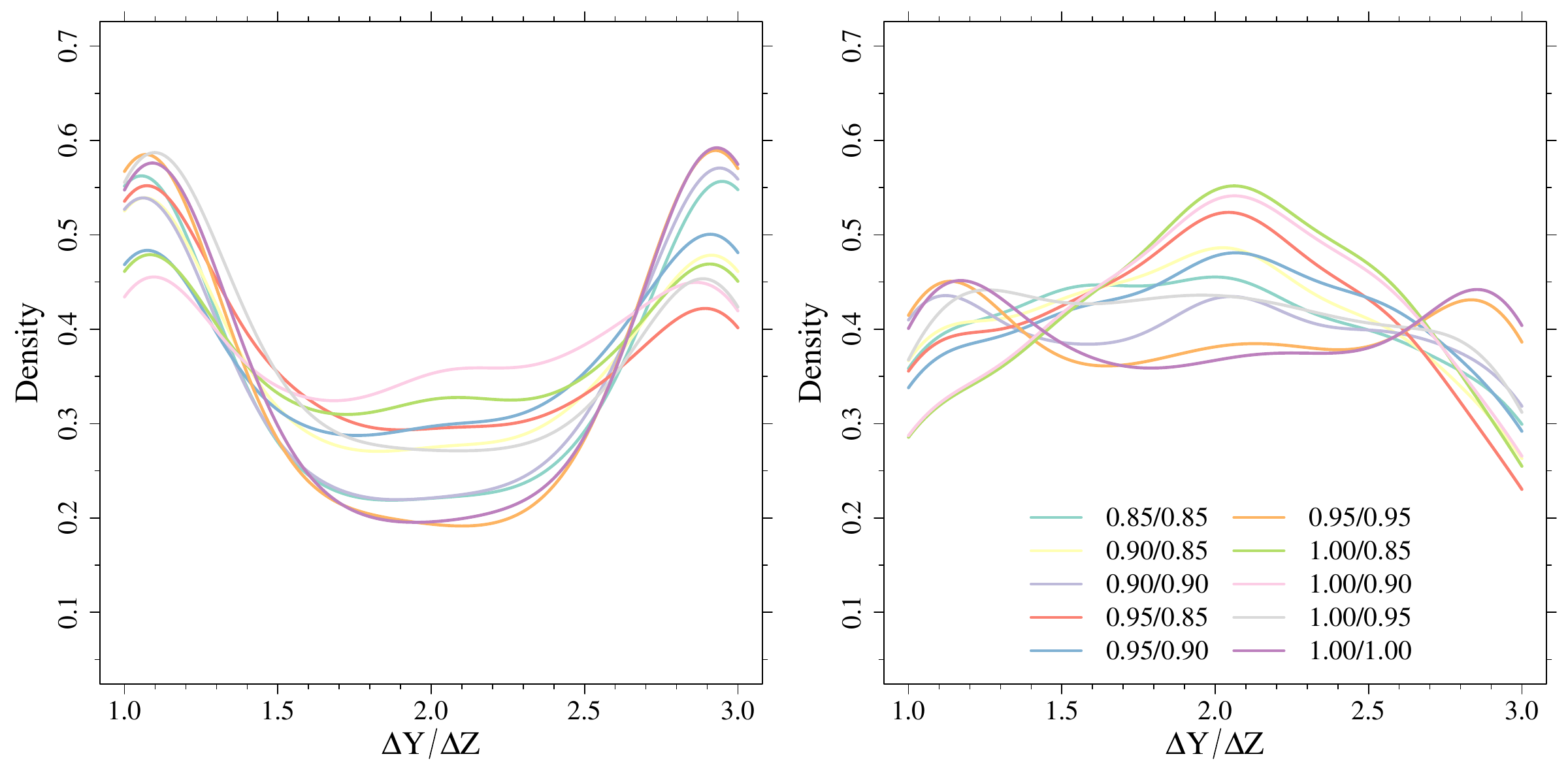} 
	\caption{Kernel density estimators of the helium-to-metal enrichment ratio for different binary systems. {\it Left}: Density estimates obtained employing error set S1.
	{\it Right}: Same as in the Left panel but adopting error set S2.	     
	}
	\label{fig:dydz}
\end{figure*}

\begin{figure*}
	\centering
	\includegraphics[width=17.5cm,angle=0]{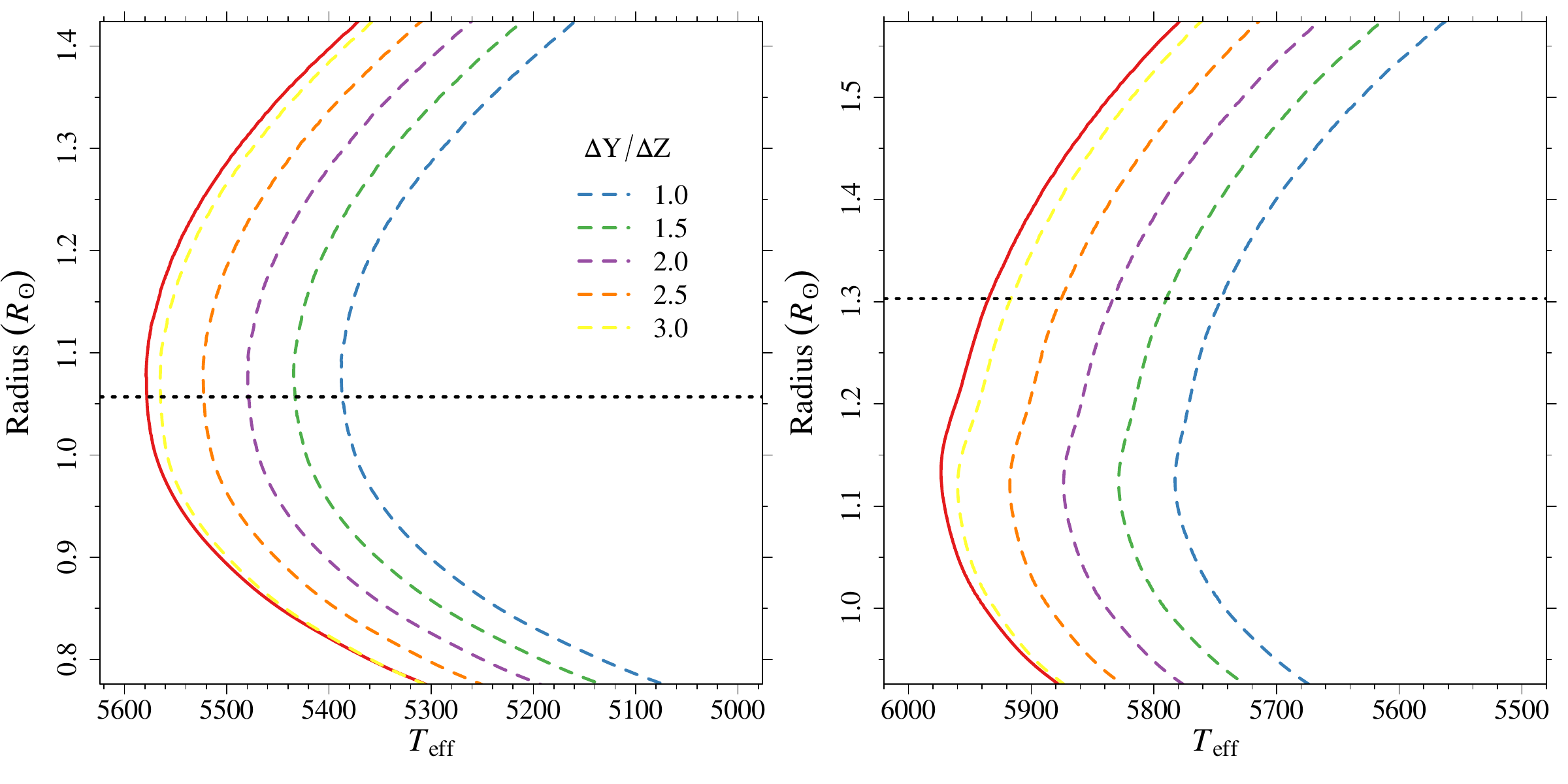} 
	\caption{Stellar tracks for different masses, [Fe/H], and $\Delta Y/\Delta Z$ {\it Left}: Evolutionary tracks for $M = 0.85$ $M_{\sun}$. The solid line corresponds to [Fe/H] = 0.0, and $\Delta Y/\Delta Z$ = 2.0. Dashed lines correspond to [Fe/H] = 0.1, for different values of  $\Delta Y/\Delta Z$. The dotted line marks the radius value at 80\% of the evolution in MS.
		{\it Right}: Same as in the Left panel but for $M = 1.00$ $M_{\sun}$.
	}
	\label{fig:tracks}
\end{figure*}

\begin{figure*}
	\centering
	\includegraphics[width=17.5cm,angle=0]{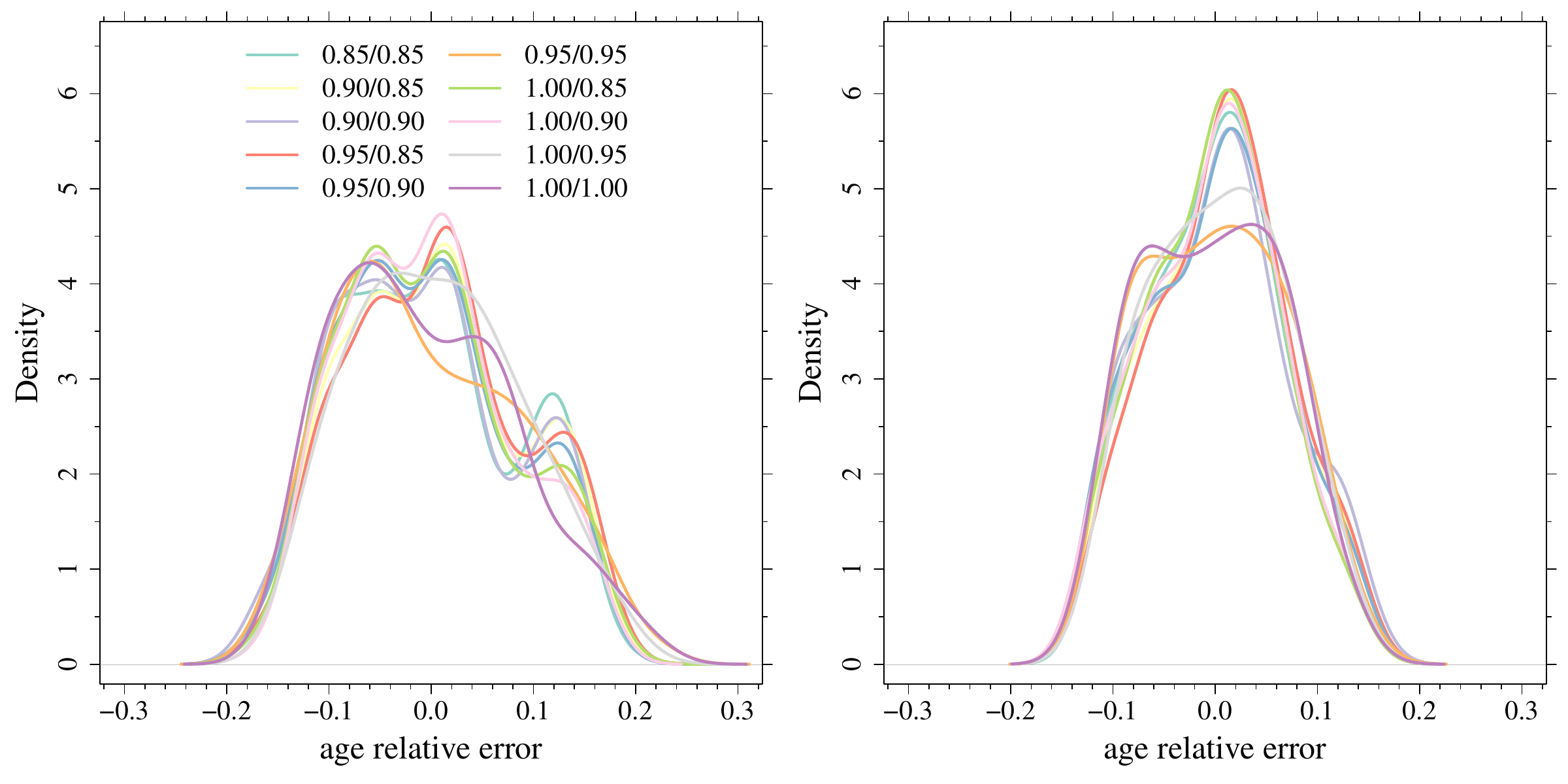} 
	\caption{Kernel density estimators of the relative error in the estimated age for different binary systems. {\it Left}: Densities obtained adopting error set S1.
		{\it Right}: Same as in the Left panel but using errors set S2.	     
	}
	\label{fig:age}
\end{figure*}

The Monte Carlo simulations reveal that the estimation of the $\Delta Y/\Delta Z$ parameter is extremely challenging in all examined scenarios.
The left panel of Figure~\ref{fig:dydz} shows the posterior distributions
of the helium-to-metal enrichment ratio for the studied binary systems assuming S1 uncertainties.
All distributions display a similar pattern, 
with two peaks located at the edges of the explored range rather than close to the true value of $\Delta Y/\Delta Z = 2.0$. 
The situation slightly improves  when considering the S2 lower observational uncertainties (right panel in Fig.~\ref{fig:dydz}), because some systems show a posterior mildly peaked around the true value. 
However all the posterior distributions have a large variance and the $\Delta Y/\Delta Z$ parameter is effectively unconstrained across the entire range.
Therefore, even in this over-optimistic scenario in which artificial stars and models perfectly align save for random uncertainties, binary systems fail to provide a reliable estimate of the helium-to-metal enrichment ratio. 
The estimation is even more problematic when attempted using real-world binary systems, as systematic discrepancies between models and stars are expected.

The presence of peaks at the edges of the considered range is not surprising, as it has been reported in previous works both using artificial and real binary systems \citep{TZFor, BinTeo, Valle2023b}. 
The underlying cause of these peaks and the differences between the two panels in Figure~\ref{fig:dydz} lies in the interplay between helium entrenchment and the effect of [Fe/H] on the effective temperature.
The two panels of Fig.~\ref{fig:tracks} show  the stellar evolutionary tracks for a 0.85 $M_{\sun}$ star (left panel) and 1.00  $M_{\sun}$ star (right panel), both with [Fe/H] = 0.0 and $\Delta Y/\Delta Z = 2.0$, alongside other tracks for the same masses, but with [Fe/H] = 0.1 and varying  $\Delta Y/\Delta Z$. 
A metallicity error of 0.1 dex, equivalent to the assumed observational uncertainty in S1 scenario, displaces the stellar track of 0.85 $M_{\sun}$ model in effective temperature from 5578 K to 5478 K. This difference cannot be compensated even by the most helium-rich model, because its effective temperature at the radius constraint is only 5564 K. Consequently, in this scenario the likelihood algorithm only considers models with $\Delta Y/\Delta Z = 3.0$ in the fit. A similar pattern is observed for 1.00 $M_{\sun}$ models.
An opposite scenario arises when [Fe/H] is underestimated, leading to an over representation of solutions with  $\Delta Y/\Delta Z = 1.0$. Only when the perturbations in [Fe/H] and $T_{\rm eff}$ counteract each other  are solutions within the middle of the range explored.
The presence of the two significant peaks at the edge of the range is therefore an artefact. If a broader range of $\Delta Y/\Delta Z$ is allowed 
during the fitting process, the height of the peaks in the posterior distributions diminish and the distributions tend to become nearly uniform, as observed for several systems in the right panel of the figure.

\begin{table}[ht]
	\centering
	\caption{Percentiles of the relative error in the age estimates from different binary systems.}\label{tab:age}
	\begin{tabular}{lcccccc}
		\hline\hline
		 & \multicolumn{3}{c}{S1} & \multicolumn{3}{c}{S2}\\
		 Masses ($M_{\sun}$) & $q_{16}$ & $q_{50}$ & $q_{84}$ & $q_{16}$ & $q_{50}$ & $q_{84}$ \\ 
		\hline
		 0.85/0.85 & -9.7 & -0.2 & 11.1 & -6.8 & 0.3 & 6.0 \\ 
		 0.90/0.85 & -9.6 & 0.3 & 11.3 & -6.5 & 0.6 & 6.3 \\ 
		 0.90/0.90 & -9.7 & -0.4 & 11.4 & -8.0 & 0.5 & 7.2 \\ 
		 0.95/0.85 & -9.0 & 0.8 & 11.3 & -6.0 & 0.8 & 6.6 \\ 
		 0.95/0.90 & -9.9 & -0.3 & 9.8 & -6.8 & 0.3 & 6.5 \\ 
		 0.95/0.95 & -9.4 & -1.2 & 10.4 & -7.8 & 0.0 & 7.4 \\ 
		 1.00/0.85 & -9.7 & -0.2 & 9.1 & -6.5 & 0.2 & 6.2 \\ 
		 1.00/0.90 & -9.0 & -0.2 & 8.5 & -6.7 & 0.2 & 6.3 \\ 
		 1.00/0.95 & -8.2 & 0.4 & 9.5 & -7.0 & 0.4 & 7.3 \\ 
		 1.00/1.00 & -9.8 & -1.7 & 8.4 & -7.9 & -0.2 & 7.3 \\ 
		\hline
	\end{tabular}
	\tablefoot{In the first column, the masses of the primary and secondary stars are provided. Columns 2 to 4 contain the percentiles of the relative error in the estimated ages under uncertainty set S1. Columns 5 to 8 present the corresponding percentiles for uncertainty set S2. } 
\end{table}

The difficulty in accurately determining the $\Delta Y/\Delta Z$ value has a limited influence on the estimation of the systems age.
Figure~\ref{fig:age} shows that the relative error in the age remains centred around zero even under scenario S1. Table~\ref{tab:age} reports the median values of these distributions ($q_{50}$) and the corresponding 16-th and 84-th percentiles ($q_{16}$ and $q_{84}$ respectively), adopted as the boundary of the 1 $\sigma$ interval. The average width of these distributions, calculated as the half-width of the interval between 
$q_{16}$ and $q_{84}$, is about 10\%.
When observational data are more precise, the distributions of the age relative error shrink (right panel of Fig.~\ref{fig:age} and last three columns in Table~\ref{tab:age}). In fact, the average width of these distributions
is about 6.5\% that is, three-quarters of the values found in S1 scenario.
In the S1 scenario 
the marginalized posterior distribution of the relative errors in stellar ages exhibits a multi-modal feature, which arises from the presence of multiple peaks in the estimated helium-to-metal enrichment ratios.
This feature is less prominent in the S2 scenario, as evidenced by the right panel of Figure~\ref{fig:age}. Several systems in this scenario exhibit a  unimodal distribution.

\subsection{Artificial stars with [Fe/H] = $-0.3$}

\begin{figure*}
	\centering
	\includegraphics[width=17.5cm,angle=0]{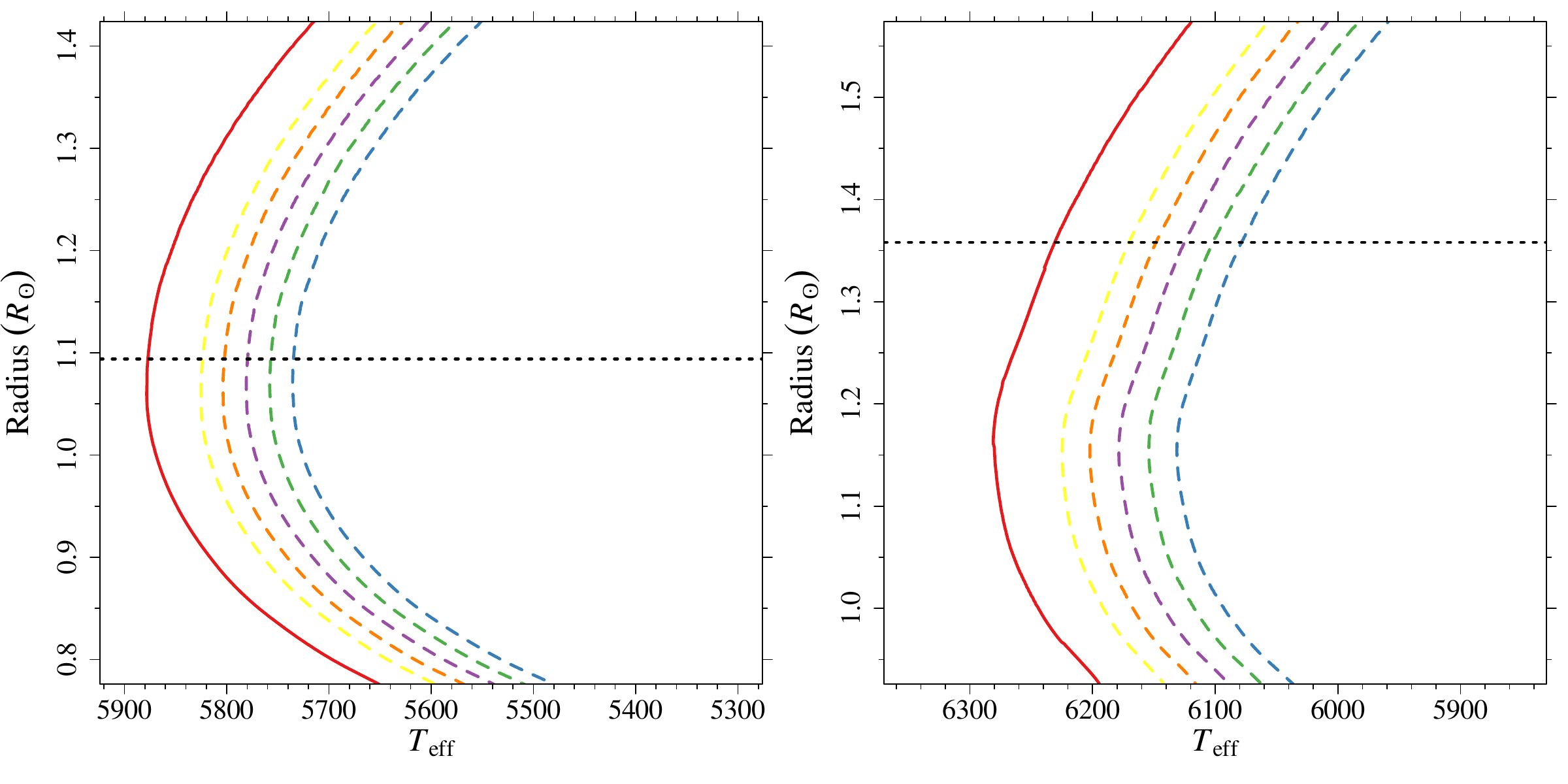} 
	\caption{As in Fig.~\ref{fig:tracks}, but assuming [Fe/H] = $-0.3$ as reference. Dashed tracks have [Fe/H] = $-0.2$.
	}
	\label{fig:tracks-03}
\end{figure*}

\begin{figure*}
	\centering
	\includegraphics[width=17.5cm,angle=0]{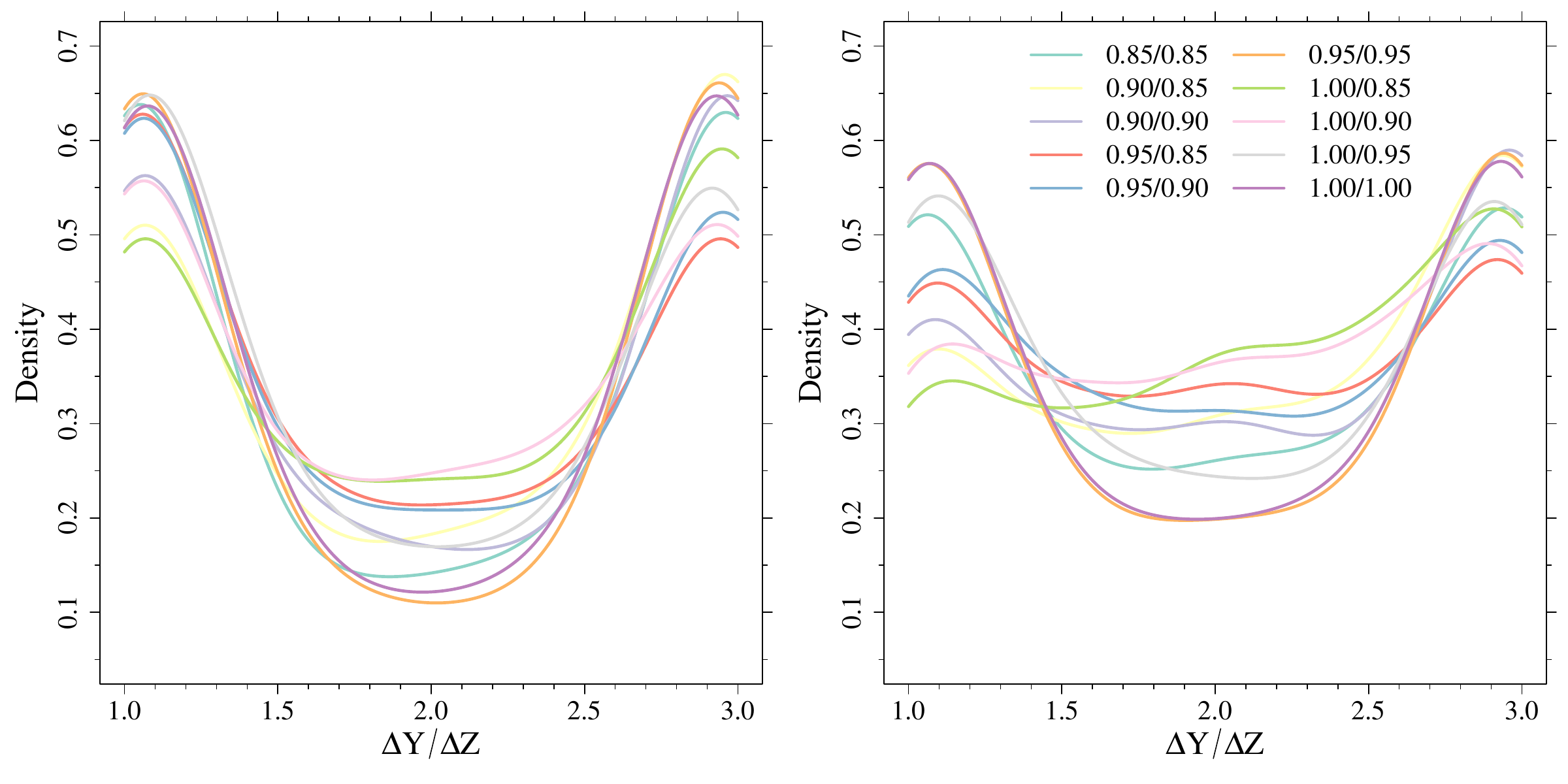} 
	\caption{As in Fig.~\ref{fig:dydz}, but assuming [Fe/H] = $-0.3$ as reference.    
	}
	\label{fig:dydz-03}
\end{figure*}

Results from Monte Carlo simulations performed for artificial stars with a baseline metallicity of [Fe/H] = $-0.3$ are even less encouraging than those discussed in the previous section. This unsurprising finding stems from the fact that the lower the metallicity [Fe/H], the narrower the range $\Delta Y$ spanned by the models. As a consequence, the range in effective temperature spanned by a set of models at fixed [Fe/H] and varying $\Delta Y/\Delta Z$ shrinks. This is clearly seen from Fig.~\ref{fig:tracks-03} which shows that the extreme variation in the $\Delta Y/\Delta Z$ parameter can compensate about one half of the $T_{\rm eff}$ offset due to a variation of 0.1 dex in the metallicity [Fe/H]. Consequently, the posterior distributions of  $\Delta Y/\Delta Z$ are even more peaked at the edges of the allowed range. 
Even in the S2 scenario, the estimated helium-to-metal enrichment ratio is significantly distorted at this baseline metallicity, as it is shown in the right panel of Fig.~\ref{fig:dydz-03}.

\section{Conclusions}\label{sec:conclusions}

We investigated the feasibility of accurately determining the helium-to-metal enrichment ratio $\Delta Y/\Delta Z$ from precise observations of binary systems.  To this end, we performed a theoretical Monte Carlo investigation. Synthetic systems were sampled from a grid of computed stellar models, and the same grid was  utilized in the subsequent fitting process.  Observational errors were accounted for by adding to the artificial system Gaussian noise. 
Two distinct scenarios were examined: S1 corresponding to realistically attainable precision of 100 K in effective temperature and 0.1 dex in surface metallicity [Fe/H], and S2 in which these uncertainties were halved.

Synthetic binary systems were constructed by pairing primary stars with masses between 0.85 and 1.00 $M_{\sun}$  with secondary stars in the same mass range. Two baseline metallicities were considered: [Fe/H] = 0.0 and $-0.3$. 
The systems were drawn from a reference grid with  $\Delta Y/\Delta Z = 2.0$ at the point when the primary star reaches 80\% of its MS evolution.
This near turn-off phase is particularly sensitive to the influence of varying helium-to-metal enrichment ratios  \citep{binary}.
From each system $N = 5,000$ perturbed Monte Carlo realization were generated.
These synthetic systems were subjected to fitting procedures to determine their age, metallicity $Z$, and initial helium abundance $Y$. The resulting best-fit  $\Delta Y/\Delta Z$ were computed. 
 
Examination of the posterior distributions of the helium-to-metal enrichment ratio revealed significant biases in the fitted parameter. 
The distributions were severely biased towards the edge of the allowable range when employing the S1 errors scenario. The situation only marginally improved when considering the S2 scenario. This behaviour arises from the magnitude of the impact of varying  $\Delta Y/\Delta Z$ on the effective temperatures of stellar models.
The considered metallicity uncertainty of 0.1 dex in the S1 scenario produces a larger shift in effective temperature compared to that caused by the extreme variations in helium-to-metal enrichment ratio. This leads to an over-abundance of edge values in the $\Delta Y/\Delta Z$  posterior distributions. 
This effect is more pronounced at lower metallicities, as the same variability in $\Delta Y/\Delta Z$ translates in a reduced variation in initial helium abundance $Y$. 
Therefore, the results presented in this paper are not dependent on the specific fitting algorithm employed and have general relevance.

While the mass range and the evolutionary phase were carefully selected to minimize or eliminate certain sources of uncertainty, such as the need to account for convective core overshooting efficiency, our findings indicate that the random variability in effective temperature and metallicity introduced by observational uncertainties significantly hinders the accurate determination of the $\Delta Y/\Delta Z$ parameter.
This result confirm the finding of \citet{BinTeo} 
who reported a similar pattern in the posterior distributions of the $\Delta Y/\Delta Z$ parameter in their theoretical study of more massive and evolved binary systems.
The inability to accurately retrieve the true value of the helium-to-metal enrichment ratio, as discussed above, emerges from a theoretical setting devoid of systematic discrepancies between synthetic systems and the underlying model grid. When employing real-world binary systems for this task, even more substantial biases are anticipated. 
Ultimately, the adoption of binary systems for  $\Delta Y/\Delta Z$ investigation 
appears to be questionable.
Fortunately, while determining the precise value of $\Delta Y/\Delta Z$ can be challenging, this has minimal impact on the system age estimation, which achieved a 10\% median precision across all studied configurations.

\begin{acknowledgements}
We thank our anonymous referee for the useful comments and suggestions.
G.V., P.G.P.M. and S.D. acknowledge INFN (Iniziativa specifica TAsP) and support from PRIN MIUR2022 Progetto "CHRONOS" (PI: S. Cassisi) finanziato dall'Unione Europea - Next Generation EU.
\end{acknowledgements}

\bibliographystyle{aa}
\bibliography{biblio}

\appendix

\section{Correlations among binary system parameters}\label{app:corr}

This appendix presents the average correlation matrix (Table~\ref{tab:corr}) for the eight stellar system parameters. The correlations imposed during the Monte Carlo simulations between the two stellar effective temperatures and the two metallicities are recovered in the fitted parameters (Kendall's $\tau$ correlation coefficient 0.97 and 0.92, respectively). The correlation between stellar masses is weaker than imposed in the simulations ($\tau=  0.58$).
Additionally, the fitted parameters exhibit a significant negative correlation between effective temperature and metallicity. This is likely due to the inherent anti-correlation between these parameters in stellar evolution tracks.

\begin{table}[!h]
	\centering
	\caption{Correlation matrix for the fitted parameters, averaged over the considered systems.}\label{tab:corr}
	\begin{tabular}{lcccccccc}
		\hline\hline
       & $T_{\rm eff,1}$ & [Fe/H]$_1$  & $R_1$ & $M_1$ & $T_{\rm eff,2}$ & [Fe/H]$_2$  & $R_2$ & $M_2$\\
       \hline
$T_{\rm eff,1}$ & 1.00 & & & &  && & \\
${\rm [Fe/H]}_1$   &-0.43 & 1.00 & &  &  &  & &\\
$R_1$     &-0.14 &-0.08 & 1.00  &&&&&\\
$M_1$     & 0.00 & 0.00 & 0.00 & 1.00   &&&&\\
$T_{\rm eff,2}$ & 0.97 &-0.43 &-0.13 & 0.00 &  1.00 &&&\\
${\rm [Fe/H]}_2$   &-0.37 & 0.92 &-0.10 & 0.00 & -0.36 & 1.00 & &\\
$R_2$     & 0.19 &-0.01 & 0.21 &-0.01 &  0.21 & 0.05 & 1.00 &\\
$M_2$     & 0.00 &-0.01 & 0.01 & 0.58 &  0.00 &-0.01 &-0.01 & 1.00\\
		\hline
	\end{tabular}
\end{table}

\end{document}